\documentclass{article}
\usepackage{spconf,amsmath,amssymb,graphicx,hyperref,array}
\usepackage{xspace}
\usepackage{enumitem}
\usepackage{microtype}
\usepackage{colortbl}
\usepackage{tabularx}
\usepackage{multirow}
\usepackage{xcolor}
\definecolor{fillzero}{HTML}{EF7B45}
\definecolor{fillmean}{HTML}{5EB1BF}
\definecolor{fillgauss}{HTML}{918EF4}
\definecolor{verdigris}{HTML}{1A7A4A}
\definecolor{spaceindigo}{HTML}{D7263D}
\usepackage{rotating}
\usepackage{booktabs}
\usepackage{makecell}
\usepackage{algorithm}
\usepackage{algorithmic}
\usepackage{balance}
\usepackage{stfloats}
\usepackage{pgfplots}
\usepackage[bottom]{footmisc}
\pgfplotsset{compat=1.18}
\usetikzlibrary{decorations.pathreplacing}
\makeatletter
\renewcommand\section{\@startsection{section}{1}{\z@}
  {-2ex plus -0.5ex minus -.2ex}
  {2ex plus .2ex}
  {\normalfont\bf\centering}}
\renewcommand\subsection{\@startsection{subsection}{2}{\z@}
  {-1.5ex plus -0.5ex minus -.2ex}
  {1.5ex plus .2ex}
  {\normalfont\bf}}
\makeatother

\newcommand{\pannssa}{\textsc{panns+sa}\xspace}
\newcommand{\pannsns}{\textsc{panns\nobreakdash-sa}\xspace}
\newcommand{\astsa}{\textsc{ast+sa}\xspace}
\newcommand{\specaug}{SpecAugment\xspace}
\newcommand{\fillzero}{\textsc{zero}\xspace}
\newcommand{\fillmean}{\textsc{mean}\xspace}
\newcommand{\fillgauss}{\textsc{gaussian noise}\xspace}

\title{Mask-Induced Displacement in Audio XAI via Logit Trajectory Decomposition}
\name{Nico Garc\'ia-Peguinho,$^1$ David Kelly,$^2$ Fabrizio Smeraldi,$^1$ Anna Xamb\'o Sed\'o$^1$}
\address{$^1$School of Electronic Engineering and Computer Science, Queen Mary University of London\\
$^2$Department of Informatics, King's College London}
\makeatletter
\long\def\@makecaption#1#2{%
  \vskip 2pt
  \setbox\@tempboxa\hbox{#1. #2}%
  \ifdim \wd\@tempboxa >\hsize #1. #2\par \else \hbox
  to\hsize{\hfil\box\@tempboxa\hfil}\fi}
\makeatother

\begin{document}
\ninept
\setlength{\textfloatsep}{8pt plus 2pt minus 2pt}
\setlength{\abovecaptionskip}{4pt}
\setlength{\belowcaptionskip}{2pt}
\maketitle
\begin{abstract}
Perturbation-based XAI methods for audio classifiers often estimate feature importance by masking spectrogram regions and crediting output changes to the retained signal. Yet they typically assume the fill (the mask replacement) is negligible. We propose logit-space trajectory decomposition to examine this assumption. An on-axis component captures output along a line connecting the fully filled (occluded) spectrogram to the fully retained original; an off-axis component captures perpendicular displacement. We evaluate across three fills, three audio classifiers, and $1{,}100$ AudioSet clips. We demonstrate that no fill is acoustically neutral under full occlusion: \fillzero activates silence and \fillgauss activates broadband noise. Under partial masking, 41--77\% of output displacement is off-axis, with the direction of the residual stable across mask retention fraction and specific to each model--fill combination. Attribution methods are unevenly exposed to off-axis displacement through their sampling and weighting strategies, revealing apparatus-dependence. Where the off-axis residual is stable and low-dimensional, its structure affords mitigation.\looseness=-1
\end{abstract}
\section{Introduction}
\label{sec:intro}

Perturbation-based e\underline{X}plainable \underline{AI} (XAI) methods explain neural classifier predictions by masking input features and crediting output changes to the retained signal. However, masking cannot remove input features, only substitute them with a replacement signal, referred to herein as \emph{fill}; every output reflects the classifier's joint response to both, introducing missingness bias~\cite{Jain2022} and risking out-of-distribution model responses~\cite{Qiu2022}. We ask: \emph{how predictably does model output change given the mixture of retained and fill signal, and what structure underlies any such departure?}\looseness=-1

The natural approach to evaluating fill choice is ablation: test various fills, measure attribution faithfulness, and select the most faithful. However, without ground-truth explanations against which to validate, evaluation methods have been shown to be circular: faithfulness evaluation rewards alignment between the fill used in attribution and evaluation, hyperparameters can produce any desired ranking~\cite{Wickstrom2024}, and models may infer class from mask shape during pixel-flipping~\cite{Rong2022}, causing persistent feature attribution disagreement~\cite{Neely2021,Krishna2022}. Conclusions drawn from perturbation-based XAI are apparatus-dependent: what is measured reflects methodological choices as much as model behaviour. We therefore forgo faithfulness evaluation in favour of a geometric investigation of model output under partial masking.\looseness=-1

\begin{figure}[t]
    \centering
    \includegraphics[width=\columnwidth]{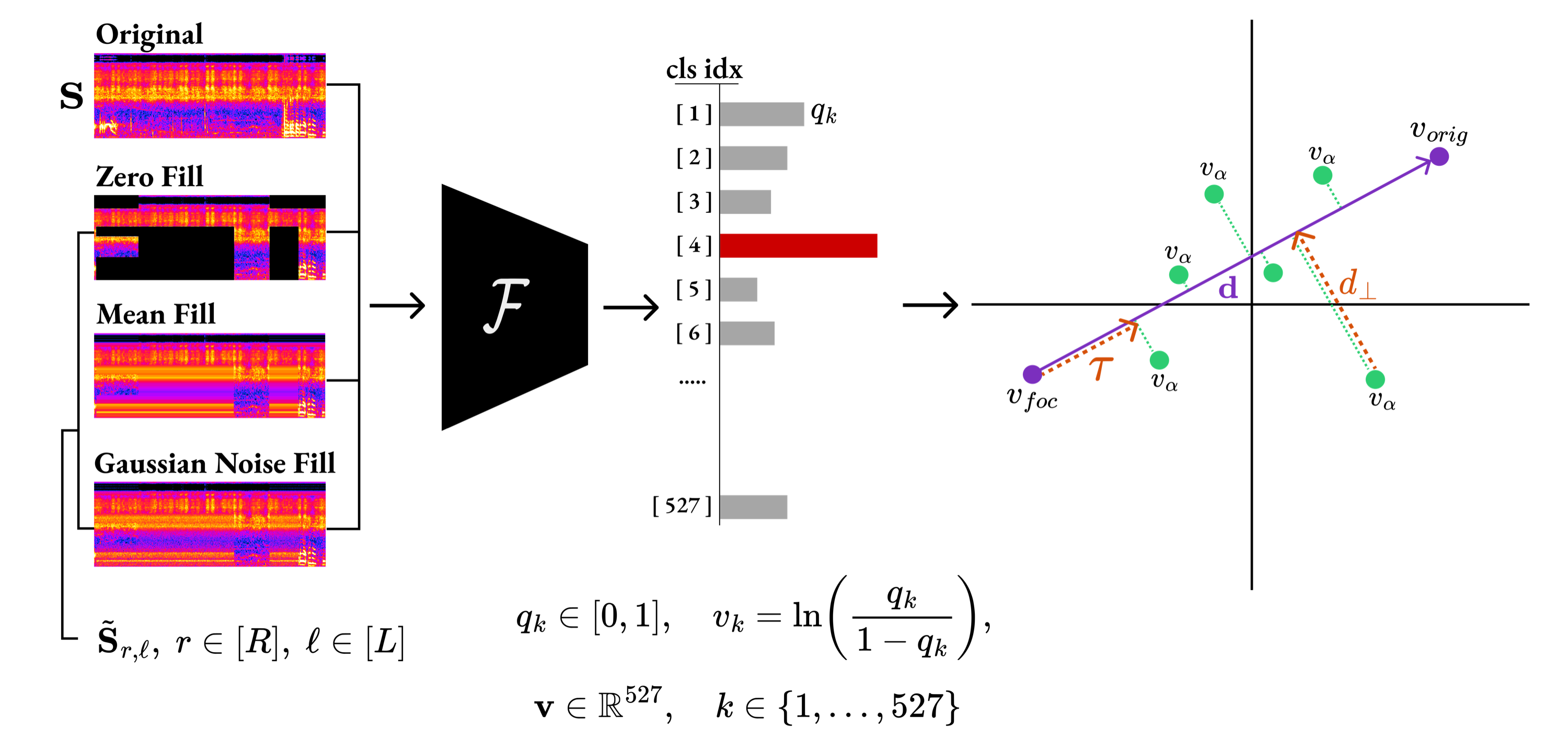}
    \vspace{-10pt}
    \caption{Method overview. Four spectrograms (original and three fills) are passed through classifier $\mathcal{F}$, producing logit vectors $\mathbf{v} \in \mathbb{R}^{527}$. Each $\mathbf{v}_\alpha$ denotes the logit vector at retention fraction $\alpha$. The logit-space trajectory is decomposed into a scalar on-axis component $\tau$ (along the attribution axis $\mathbf{d} = \mathbf{v}_\text{orig} - \mathbf{v}_\text{foc}$) and a scalar off-axis magnitude $d_\perp$, whose direction is captured by the residual vector $\mathbf{v}_\perp \in \mathbb{R}^{527}$.}
    \label{fig:method}
\end{figure}

We measure model output holistically in logit space, establishing a reference axis from the fully occluded condition (FOC, $\alpha=0$) to the original output ($\alpha=1$) as a geometric reference for decomposition. Displacement along this axis is on-axis; perpendicular displacement is off-axis. Logit trajectory decomposition characterises both components across three fills (\fillzero, \fillmean, \fillgauss), three classifiers (\astsa, \pannsns, \pannssa), and $1{,}100$ AudioSet clips. We make the following contributions:\looseness=-1
\vspace{0.5em}

\begin{enumerate}[nosep]
    \item \textbf{Partial Mask Analysis:} We propose logit-space trajectory decomposition, showing that 41--77\% of partial-mask output displacement is off-axis and invisible to scalar attribution methods (Section~\ref{sec:4.2}). This displacement concentrates in stable low-dimensional subspaces specific to each model--fill combination, with consequences for how mask sampling and surrogate weighting interact with fill-specific geometry (Section~\ref{sec:4.3}).\looseness=-1
    \item \textbf{Fully Occluded Condition:} We show no fill constitutes an information-absent baseline, with each fill activating a distinct AudioSet class profile under full occlusion (Section~\ref{sec:4.1}).\looseness=-1
\end{enumerate}
\vspace{0.5em}

All code and trained models are available at: \url{https://github.com/ne3k0/partial_mask_geometry_xai}

\section{Related Work}
\label{sec:lit}
Perturbation-based XAI methods, such as KernelSHAP~\cite{Lundberg2017}, LIME~\cite{Ribeiro2016}, and RISE~\cite{Petsiuk2018}, have seen comparatively little study in the audio domain relative to vision~\cite{Akman2024}, despite widespread audio classifier deployment~\cite{Parker2025}. Nonetheless, they have been applied across audio tasks, from singing voice detection~\cite{Mishra2020} and music tagging~\cite{Haunschmid2020} to automatic speech recognition~\cite{Wu2023}, audio content analysis~\cite{Marinelli2024}, and speech emotion recognition~\cite{Hjuler2025}. Perturbation methods have also been used to generate causal explanations~\cite{Kelly2026a}, revealing model-specific feature reliance in audio classifiers~\cite{Kelly2026b}, first noted by Sturm's Clever Hans critique~\cite{Sturm2014} and confirmed unresolved a decade later~\cite{Green2024}. Learned-surrogate approaches train decoder networks on internal classifier representations~\cite{Parekh2022,Paissan2024}, with faithfulness gains at the cost of surrogate interpretability.\looseness=-1 

Fill non-neutrality has been explored in feature masking literature~\cite{Fong2017}, including work framed as feature removal~\cite{Covert2021}, where fill dominates faithfulness rankings~\cite{Blucher2024}, and in integrated gradients, where baseline choice similarly shapes attribution~\cite{Sturmfels2020}. Black-box treatment and interpretability motivate perturbation-based methods. Core differences between audio and visual modalities, namely that mel spectrograms encode non-isotropic frequency-temporal information and \mbox{\specaug}~\cite{Park2019} applies zero-fill temporal and spectral masking during training, mean that approaches inherited from vision require dedicated study in the audio domain.\looseness=-1

\section{Method}
\label{sec:method}

\subsection{Models \& Experiment Setup}

We evaluate three mel spectrogram audio classifiers trained on AudioSet~\cite{Gemmeke2017}: \pannssa~\cite{Kong2020}, \pannsns~\cite{Kong2020}, and \astsa~\cite{Gong2021}. The two PANNs variants share the same training data, isolating \specaug as the sole training difference. \astsa is evaluated with \specaug only, as omitting it causes overfitting and notable performance degradation. Our perturbation pipeline replicates each model's signal chain exactly.\looseness=-1

The evaluation dataset is drawn from the AudioSet eval split. Fifty clips are sampled from 22 classes\footnote{Bagpipes, Boing, Chicken/rooster, Didgeridoo, Dog, Drum kit, Frog, Frying food, Gunshot/gunfire, Hair dryer, Harmonica, Heart sounds/heartbeat, Insect, Owl, Rain, Rub, Sewing machine, Speech, Thunder, Timpani, Train, Whispering.}, yielding $1{,}100$ clips, applied to each model and fill condition. Classes were selected to span diverse acoustic signal types, not sampled at random from AudioSet. Model output is always measured against the eval clip's ground truth (GT) label, not the model's top-1 prediction.\looseness=-1

\subsection{Spectrogram Segmentation, Sampling \& Perturbation}

\noindent\textbf{STFT-domain Perturbation:} Fill is applied in complex STFT space: 
\begin{equation}
    \tilde{\mathbf{Z}} = \mathbf{m} \odot \mathbf{Z} + (1 - \mathbf{m}) \odot \mathbf{Z}_{\text{fill}},
\end{equation}
where $\odot$ denotes the Hadamard product, $\mathbf{m} \in \{0,1\}^{F \times T}$ is a binary retention mask, and $\mathbf{Z}_{\text{fill}}$ the fill signal. Each model's signal chain $\Psi: \mathbb{C}^{F \times T} \to \mathbb{R}^{M \times T}$ maps $\tilde{\mathbf{Z}}$ to the log mel spectrogram $\tilde{\mathbf{S}}$ the classifier receives, discarding phase information.
\vspace*{0.5em}

\noindent\textbf{Segmentation and Sampling:} Masks are defined over a coarser $M \times X$ cell grid, with time boundaries at multiples of $T/10$ (each cell ${\approx}100$\,ms), then upsampled to STFT resolution: frequency bins via mel filterbank support, time frames via precomputed hard assignment to their enclosing coarse window, yielding a binary mask $\mathbf{m} \in \{0,1\}^{F \times T}$. The grid is partitioned into $K = K_F \times K_T = 25$ contiguous rectangular blocks with boundaries redrawn each seed, following the coalition-sampling strategy of RISE~\cite{Petsiuk2018}, where each block constitutes a coalition of cells sampled together. With $R = 300$ seeds and $L = 32$ samples per seed, each clip yields $9{,}600$ partial masks; $\alpha \in [0,1]$ is the fraction of cells retained per partial mask, binomially distributed around $0.5$ (Figure~\ref{fig:logit_traj}).\looseness=-1

We adopt RISE-style sampling because redrawing boundaries each seed ensures any cell appears across many coalitions, avoiding structural assumptions about co-occurring regions, and, over sufficiently many samples and configurations, minimising positional bias from cells differing in how strongly they drive the model output. With fixed grids, cell membership within each block is permanent, so output variation at a given $\alpha$ partly reflects which blocks happen to contain important cells rather than $\alpha$ alone, limiting any test of LIME and KernelSHAP's assumption that $\alpha$ proxies proximity to the original signal. Masks are deterministic, enabling direct comparison of fill effects across conditions.\looseness=-1
\vspace*{0.5em}

\noindent Let $Z_{f,t}$ be the element of $\mathbf{Z}$ at bin $f$, frame $t$, with magnitude $|Z_{f,t}|$ and phase $\theta_{f,t}$; $\mu_f = \frac{1}{T}\sum_{t=1}^T |Z_{f,t}|$ is the time-averaged magnitude for each frequency bin. The fill elements $Z_{\text{fill},f,t}$ are defined by one of three strategies:

\begin{enumerate}[itemsep=2pt,topsep=2pt]
    \item \textbf{\fillzero:} $Z_{\text{fill},f,t} = 0$
    \item \textbf{\fillmean:} $Z_{\text{fill},f,t} = \mu_f\, e^{j\theta_{f,t}}$ — replaces magnitude with the time-averaged $\mu_f$.
    \item \textbf{\fillgauss:} $Z_{\text{fill},f,t} = |\epsilon_{f,t}|\, e^{j\theta_{f,t}}$, where $\epsilon_{f,t} \sim \mathcal{N}(0, \mu_f^2)$ — magnitude is half-normal, scaled per frequency bin by the clip's average spectral magnitude $\mu_f$. Drawn once per clip and held constant across all masks.
\end{enumerate}

\subsection{Logit-Space Trajectory Analysis}
\label{sec:3.3}

Model outputs are analysed in logit space: $v_k = \ln(q_k / (1 - q_k))$ for each class $k$, giving $\mathbf{v} \in \mathbb{R}^{527}$ (one dimension per AudioSet class). $\bar{q}$ denotes mean sigmoid confidence across clips. This linearises the sigmoid outputs for geometric analysis. Let $\mathbf{v}_\text{orig}$ and $\mathbf{v}_\text{foc}$ denote the logit vectors of a clip's original and fully occluded inputs respectively, and let $\mathbf{d} = \mathbf{v}_\text{orig} - \mathbf{v}_\text{foc} \in \mathbb{R}^{527}$ be the attribution axis (hereafter $\mathbf{d}$; Fig.~\ref{fig:method}). Although $\mathbf{d}$ is clip-specific, the quantities derived from it are normalised, preserving cross-clip comparability. For a masked output at retention fraction $\alpha$, let $\Delta\mathbf{v} = \mathbf{v}_\alpha - \mathbf{v}_\text{foc} \in \mathbb{R}^{527}$ denote its displacement from the fully occluded output. We decompose $\Delta\mathbf{v}$ along $\mathbf{d}$:
\begin{equation}
\tau = \frac{\Delta\mathbf{v} \cdot \mathbf{d}}{\|\mathbf{d}\|^2}, \qquad d_{\perp} = \|\Delta\mathbf{v} - \tau\,\mathbf{d}\|, \qquad d_{\text{total}} = \|\Delta\mathbf{v}\|.
\end{equation}
$\tau \in \mathbb{R}$ is the scalar projection onto $\mathbf{d}$ (so $\tau\mathbf{d}$ in the second term is a scaled vector), anchored so that $\tau = 0$ at FOC and $\tau = 1$ at the original output. $d_\perp$ is the magnitude of the displacement orthogonal to $\mathbf{d}$. The additive surrogate assumption requires $\tau \approx \alpha$. Large $d_\perp$ evidences output activation in directions the attribution axis cannot account for. The scalar $\tau_\text{gt} = \Delta v_\text{gt} / d_\text{gt}$ gives the normalised ground-truth (GT) class logit displacement by which attribution methods assign feature importance to each partial mask, where $d_\text{gt}$ is the GT-class component of $\mathbf{d}$. We report $\bar{\sigma}_{\boldsymbol{\tau}}$ and $\bar{\sigma}_{\boldsymbol{d_{\perp}}}$, the mean per-clip standard deviations of $\tau$ and $d_\perp$, and $d_{\perp}/d_{\text{total}}$, the mean off-axis fraction of total output displacement.

\subsection{Structure of the Off-Axis Residual}
\label{sec:3.4}

The residual vector $\mathbf{v}_\perp = \Delta\mathbf{v} - \tau\,\mathbf{d} \in \mathbb{R}^{527}$ is perpendicular to $\mathbf{d}$ by construction. To characterise the structure of $\mathbf{v}_\perp$ within each retention bin, we estimate its within-bin covariance. Fix a mel spectrogram classifier $\mathcal{F}$ and fill $\mathbf{Z}_{\text{fill}}$. For each retention bin $b$ (corresponding to $\alpha \in [\alpha_b, \alpha_{b+1})$, with $B = 20$ bins), pool all $\mathbf{v}_\perp$ vectors across the set of 22 evaluation classes and all clips, yielding $n_b$ vectors. Retention bins follow a binomial distribution, with minimum $n_b = 9{,}891$ across valid bins, giving a sample-to-dimension ratio of at least 19:1. Pooling across clips is intentional: it enables detection of shared off-axis structure across the clip ensemble, not just within individual clips. The within-bin covariance is:
\begin{equation}
\boldsymbol{\Sigma}_b = \mathbb{E}\!\left[\mathbf{v}_\perp \mathbf{v}_\perp^\top \mid \alpha \in b\right] - \boldsymbol{\mu}_b \boldsymbol{\mu}_b^\top \in \mathbb{R}^{527 \times 527},
\end{equation}
where $\boldsymbol{\mu}_b$ is the bin mean vector. Subtracting it centres $\boldsymbol{\Sigma}_b$; since each $\mathbf{v}_\perp$ is perpendicular to its clip-specific $\mathbf{d}$, $\boldsymbol{\mu}_b$ averages across different perpendicular subspaces and carries no consistent geometric interpretation. All estimates use a numerically stable single-pass batch update~\cite{Welford1962}. 

Eigendecomposition of $\boldsymbol{\Sigma}_b$ gives eigenvalues $\lambda_1 \geq \cdots \geq \lambda_{526} \geq 0$. We summarise the spectral concentration of each bin with the participation ratio:
\begin{equation}
\mathrm{PR}_b = \frac{\left(\sum_i \lambda_i\right)^2}{\sum_i \lambda_i^2},
\end{equation}
which gives the effective number of dimensions occupied by $\mathbf{v}_\perp$ in bin $b$. $\mathrm{PR}_b = 1$ for a rank-1 distribution concentrated along a single direction; $\mathrm{PR}_b = 526$ in the isotropic limit, since $\mathbf{v}_\perp$ is constrained orthogonal to $\mathbf{d}$ by construction. Table~\ref{tab:off_axis} reports the mean $\mathrm{PR}$ across valid bins.\looseness=-1

To test whether the dominant off-axis direction is consistent across retention fraction, we extract the leading eigenvector $\mathbf{e}_b$ of $\boldsymbol{\Sigma}_b$ for each bin and compute the pairwise absolute cosine similarity $c_{b,b'} = |\mathbf{e}_b \cdot \mathbf{e}_{b'}|$ across all bin pairs, with absolute value handling eigenvector sign ambiguity. The mean and minimum pairwise cosine similarity, denoted $\bar{c}$ and $\min c_{b,b'}$, characterise directional stability across $\alpha$. We additionally report $\bar{\rho}$, the mean over bins of the fraction of off-axis variance explained by the leading direction (Table~\ref{tab:off_axis}).

\section{Results \& Discussion}
\label{sec:results}
\subsection{Fill Non-Neutrality Under Full Occlusion}
\label{sec:4.1}

\begin{table*}[!t]
\fontsize{8}{10}\selectfont
\centering
\setlength{\tabcolsep}{6pt}
\renewcommand{\arraystretch}{1.2}
\caption{Top-2 predicted classes per fill at FOC ($N=1{,}100$ clips, 22 classes). $\bar{q}$: mean sigmoid confidence. $\sigma$ omitted for \fillzero (signal-independent).}
\vspace{0.5em}
\label{tab:foc_top2}
\begin{tabularx}{\textwidth}{@{} l l >{\raggedright\arraybackslash}X r r r >{\raggedright\arraybackslash}X r r r @{}}
\hline
Model & Fill & Top-1 class & $\bar{q}$ & $\sigma$ & $n$ & Top-2 class & $\bar{q}$ & $\sigma$ & $n$ \\
\hline
\multirow{3}{*}{\textbf{\astsa}}
 & \textcolor{fillzero}{\fillzero}   & Silence     & $.608$ & --     & 1100 & Music       & $.269$ & --     & 1100 \\
 & \textcolor{fillmean}{\fillmean}   & Mains hum   & $.691$ & $.220$ &  857 & Hum         & $.598$ & $.236$ &  864 \\
 & \textcolor{fillgauss}{\fillgauss} & White noise & $.466$ & $.145$ &  969 & Noise       & $.176$ & $.092$ &  350 \\
\hline
\multirow{3}{*}{\textbf{\pannsns}}
 & \textcolor{fillzero}{\fillzero}   & Silence     & $.327$ & --     & 1100 & Music       & $.186$ & --     & 1100 \\
 & \textcolor{fillmean}{\fillmean}   & Buzzer      & $.216$ & $.123$ &  273 & Hum         & $.173$ & $.129$ &  178 \\
 & \textcolor{fillgauss}{\fillgauss} & White noise & $.342$ & $.136$ &  618 & Static      & $.234$ & $.121$ &  545 \\
\hline
\multirow{3}{*}{\textbf{\pannssa}}
 & \textcolor{fillzero}{\fillzero}   & Music       & $.187$ & --     & 1100 & Silence     & $.035$ & --     & 1100 \\
 & \textcolor{fillmean}{\fillmean}   & Sine wave   & $.512$ & $.241$ &  295 & Siren       & $.226$ & $.164$ &  120 \\
 & \textcolor{fillgauss}{\fillgauss} & Static      & $.229$ & $.104$ &  348 & White noise & $.159$ & $.068$ &  401 \\
\hline
\end{tabularx}
\end{table*}

An information-absent baseline requires $\mathbf{v}_{\text{foc}} = \mathbf{0}$, that is, $q_k = 0.5$ for all $k$. No fill can guarantee this for a discriminatively trained classifier, and any fill carries an acoustic character to which trained models respond. \fillzero fill is signal-independent, yielding identical output distributions across all clips and classes within each model. $\mathbf{v}_\text{foc}$ therefore has zero clip-to-clip variance and $\sigma$ is not reported for \fillzero fill. \fillmean and \fillgauss fills are signal-dependent, introducing clip-level variance absent under \fillzero fill (Table~\ref{tab:foc_top2}).\looseness=-1

The top-1 class under \fillzero fill differs notably across architectures. \astsa and \pannsns both assign Silence ($\bar{q}=0.608$ and $\bar{q}=0.327$ respectively), congruent with zero fill's acoustic character, whereas \pannssa assigns Music ($\bar{q}=0.187$).\looseness=-1

\begin{table}[!t]
\small\centering
\fontsize{8}{10}\selectfont
\caption{Modal top-1 class and mean $\bar{q}$ per source class under \fillmean at FOC. Colours indicate semantic group: $^\dagger${\color{verdigris}electrical interference} (Mains hum, Hum, Buzzer), $^\S${\color{spaceindigo}tonal periodic} (Sine wave, Siren).\looseness=-1}
\vspace{0.5em}
\label{tab:foc_mean_cls}
\setlength{\tabcolsep}{3pt}
\renewcommand{\arraystretch}{1.2}
\begin{tabularx}{\linewidth}{@{} >{\raggedright\arraybackslash}p{1.4cm} >{\raggedright\arraybackslash}X >{\raggedright\arraybackslash}X >{\raggedright\arraybackslash}X @{}}
\hline
 & \textbf{\astsa} & \textbf{\pannsns} & \textbf{\pannssa} \\
\hline
Frying food & \textcolor{verdigris}{Mains hum}$^\dagger$ -- .83 & \textcolor{verdigris}{Hum}$^\dagger$ -- .07 & \textcolor{spaceindigo}{Siren}$^\S$ -- .13 \\
Drum kit    & \textcolor{verdigris}{Mains hum}$^\dagger$ -- .78 & \textcolor{verdigris}{Buzzer}$^\dagger$ -- .06 & Sound fx -- .18 \\
Owl         & \textcolor{spaceindigo}{Sine wave}$^\S$ -- .39 & \textcolor{verdigris}{Hum}$^\dagger$ -- .09 & \textcolor{spaceindigo}{Sine wave}$^\S$ -- .49 \\
Heartbeat   & \textcolor{spaceindigo}{Sine wave}$^\S$ -- .37 & Music -- .12 & \textcolor{spaceindigo}{Sine wave}$^\S$ -- .47 \\
Whispering  & \textcolor{verdigris}{Mains hum}$^\dagger$ -- .24 & \textcolor{verdigris}{Hum}$^\dagger$ -- .09 & \textcolor{spaceindigo}{Sine wave}$^\S$ -- .41 \\
Bagpipes    & \textcolor{verdigris}{Mains hum}$^\dagger$ -- .14 & \textcolor{verdigris}{Buzzer}$^\dagger$ -- .17 & \textcolor{spaceindigo}{Siren}$^\S$ -- .14 \\
\hline
\end{tabularx}

\end{table}

\fillmean fill is the most differentiated condition across architectures. \astsa shows strong top-1 consensus (Mains hum, 857/1100, $\bar{q}=0.691$). \pannsns shows weak top-1 consensus (273/1100, $\bar{q}=0.216$), distributing probability mass across a large set of classes. \pannssa concentrates on tonal-periodic classes (Sine wave, Siren) (Table~\ref{tab:foc_mean_cls}). \fillgauss fill yields the most semantically consistent responses across models. All three converge on broadband noise classes, with \astsa reaching the highest consensus (969/1100: White noise, $\bar{q}=0.466$) (Table~\ref{tab:foc_top2}).\looseness=-1

\fillmean and \fillgauss fills inherit spectral properties from the source clip, making fill-intrinsic discriminativeness and source signal structure inseparable at the clip level, a confound that propagates into partial-mask attribution at all retention fractions.\looseness=-1

\subsection{Logit-Space Trajectories Under Partial Masking}
\label{sec:4.2}

\begin{figure*}[!t]
\centering
\setlength{\abovecaptionskip}{4pt}
\caption{Logit-space trajectory analysis across 95.04M partial masks (22 classes, 50 clips, 9{,}600 masks/clip). Fill: \textcolor{fillzero}{Z}\,=\,zero, \textcolor{fillmean}{M}\,=\,mean, \textcolor{fillgauss}{G}\,=\,Gaussian. \textit{Row~1:} mask distribution and mean on-axis projection $\tau$ by $\alpha$. Solid line\,=\,full logit vector; dotted\,=\,GT class ($\tau_\text{gt}$); dashed\,=\,ideal ($\tau{=}\alpha$). \textit{Row~2:} per-fill summary table with $\bar{\sigma}_{\boldsymbol{\tau}}$, $\bar{\sigma}_{\boldsymbol{d_{\perp}}}$, $d_{\perp}/d_{\text{total}}$; mean off-axis residual $d_{\perp}$ by $\alpha$.}
\vspace{1.0em}
\label{fig:logit_traj}
\includegraphics[width=0.95\linewidth]{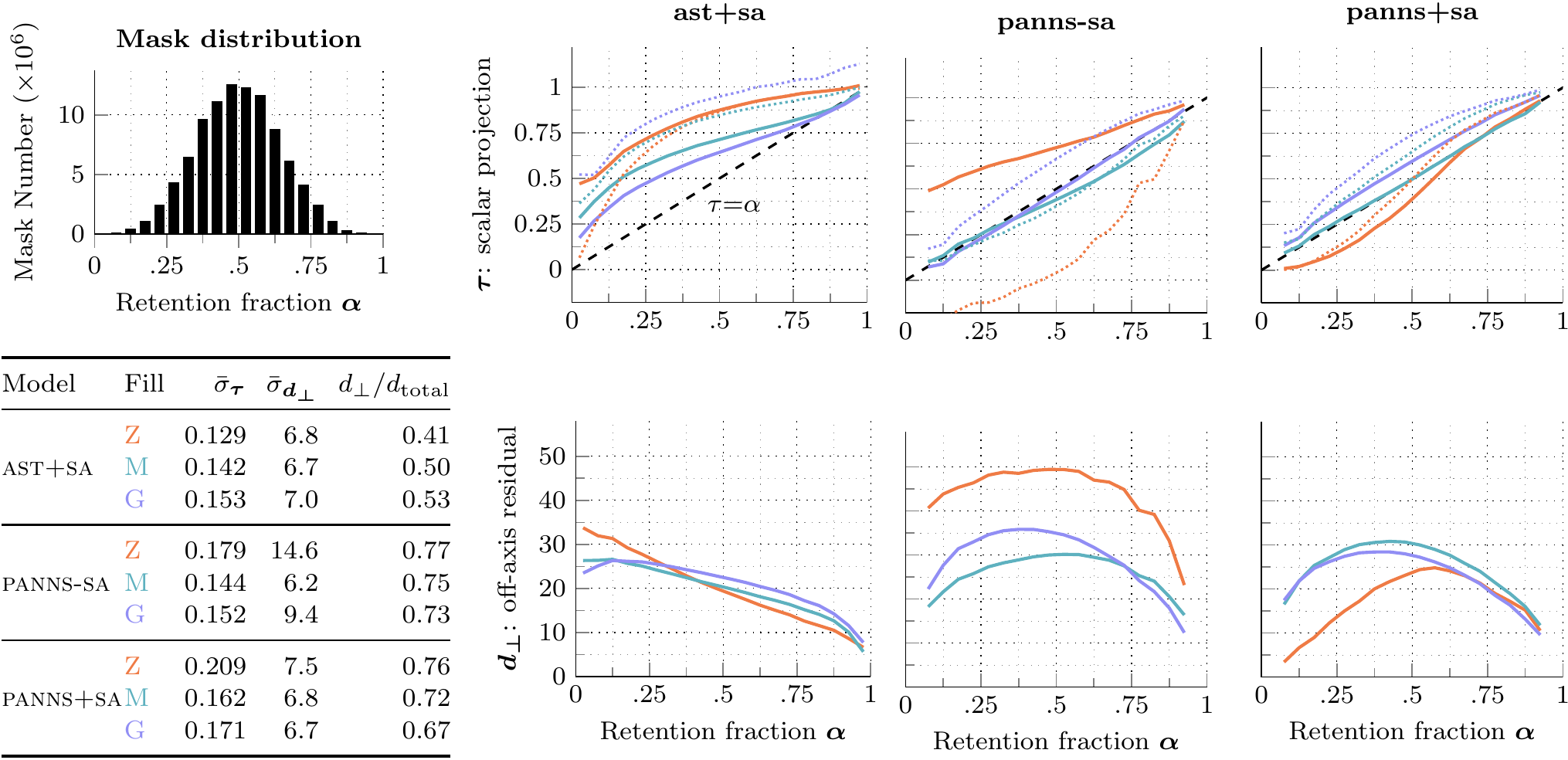}
\end{figure*}

Decomposing the output trajectory geometrically from the FOC reveals that between 41\% and 77\% of total output displacement is orthogonal to the attribution axis $\mathbf{d}$, captured by $d_{\perp}/d_{\text{total}}$ (Fig.~\ref{fig:logit_traj}). This displacement is invisible to scalar target-class methods by construction: LIME and KernelSHAP model GT-class output as a linear function of retained features and RISE tracks the scalar GT-class output, weighting all masks equally regardless of $\alpha$. Under high $d_\perp$, attribution weights are assigned in the presence of fill-driven model responses the surrogate cannot observe.\looseness=-1

\astsa is the least affected ($d_{\perp}/d_{\text{total}} \in [0.41, 0.53]$), with $d_{\perp}$ generally decreasing as $\alpha$ increases. LIME's proximity kernel, which weights masks by similarity to the original input, favours high-$\alpha$ coalitions, coinciding with the region of lowest $d_{\perp}$ for \astsa and reducing surrogate exposure to unaccounted-for off-axis activation. Under \fillgauss and \fillmean fills, both signal-dependent, $\tau_{\text{gt}}$ starts at approximately 0.52 and 0.36 at $\alpha = 0.025$ respectively, with Gaussian fill $\tau_{\text{gt}} > 1$ at $\alpha > 0.75$.\looseness=-1

Both PANNs architectures show substantially larger off-axis displacement than \astsa ($d_{\perp}/d_{\text{total}} \in [0.67, 0.77]$). \specaug reduces clip-level instability under \fillzero fill ($\bar{\sigma}_\perp$: $14.6 \to 7.5$) and dramatically suppresses zero-fill $d_\perp$ at low retention fractions, consistent with learned invariance to zero-fill masking during training. \fillmean and \fillgauss fills show minimal change across the two PANNs variants. For \pannsns, Gaussian and mean fills track the $\tau = \alpha$ reference well yet $d_{\perp}$ remains substantial. On-axis predictability and off-axis contamination coexist.\looseness=-1

Both PANNs variants exhibit an inverted-U $d_{\perp}$ profile, peaking around mid-$\alpha$ where fill and retained signal are most evenly mixed. Bernoulli block sampling with $p{=}0.5$ concentrates the majority of masks in this region by construction, precisely where off-axis displacement is highest. KernelSHAP's Shapley kernel assigns lower regression weights to mid-$\alpha$ coalitions when fitting the surrogate, coinciding with the peak of the PANNs $d_{\perp}$ profile. It is therefore better positioned to reduce the impact of PANNs off-axis behaviour. Fill shapes the logit-space trajectory in directions invisible to scalar attribution methods, with architecture and fill-specific consequences that $\tau$ alone cannot reveal.\looseness=-1

\subsection{Structure of the Off-Axis Component}
\label{sec:4.3}

\pannsns under zero fill exhibits the most structured off-axis residual (Table~\ref{tab:off_axis}): $\mathrm{PR}=8.4$ indicates that $\mathbf{v}_\perp$ occupies fewer than nine effective dimensions. $\bar{c} = 0.95$ confirms the leading direction is consistent across $\alpha$, and 32.1\% of off-axis variance is concentrated in a single eigenvector. A stable, low-dimensional off-axis profile of this kind is the most tractable target for future correction strategies.\looseness=-1

\specaug training reduces this stability markedly. \pannssa falls below $\bar{c} \leq 0.82$ across all three fills, with $\min c_{b,b'}$ as low as $0.02$ under \fillgauss, indicating that the leading off-axis direction shifts substantially across retention fraction. Direction instability across the mask ensemble makes post-hoc correction strategies less tractable.\looseness=-1

\astsa presents a distinct profile: $\bar{c}$ remains high across all three fills (0.87--0.95), with $\min c_{b,b'} \geq 0.74$ for zero and mean, indicating that the leading off-axis direction is stable across $\alpha$. However, $\mathrm{PR} \in [13.4, 19.1]$ and $\bar{\rho} \in [18.5, 23.5]\%$ show that variance is distributed across more dimensions than \pannsns under \fillzero. \astsa is directionally consistent but structurally diffuse.\looseness=-1

\begin{table}[!t]
\fontsize{8}{10}\selectfont\centering
\setlength{\tabcolsep}{3pt}
\renewcommand{\arraystretch}{1.2}
\caption{Off-axis residual structure per model--fill condition. PR: participation ratio (effective dimensionality). $\bar{c}$: mean pairwise eigenvector cosine similarity. $\min c_{b,b'}$: worst-case similarity. $\bar{\rho}$: variance fraction explained by leading eigenvector. Fill: Z\,=\,\fillzero, M\,=\,\fillmean, G\,=\,\fillgauss.\looseness=-1}
\label{tab:off_axis}
\vspace{0.5em}
\begin{tabular*}{\columnwidth}{@{\extracolsep{\fill}} l l r r r r @{}}
\hline
\textbf{Model} & \textbf{Fill} & \textbf{PR} & $\bar{c}$ & $\min c_{b,b'}$ & $\bar{\rho}$ (\%) \\
\hline
\multirow{3}{*}{\textbf{\astsa}}
  & \textcolor{fillzero}{Z}   & 19.14 & .95 & .74 & 18.5 \\
  & \textcolor{fillmean}{M}   & 13.37 & .94 & .76 & 23.5 \\
  & \textcolor{fillgauss}{G}  & 15.13 & .87 & .25 & 20.0 \\
\hline
\multirow{3}{*}{\textbf{\pannsns}}
  & \textcolor{fillzero}{Z}   & 8.39  & .95 & .68 & 32.1 \\
  & \textcolor{fillmean}{M}   & 19.56 & .86 & .20 & 15.3 \\
  & \textcolor{fillgauss}{G}  & 16.03 & .90 & .52 & 18.5 \\
\hline
\multirow{3}{*}{\textbf{\pannssa}}
  & \textcolor{fillzero}{Z}   & 17.62 & .71 & .24 & 21.6 \\
  & \textcolor{fillmean}{M}   & 16.75 & .82 & .28 & 16.7 \\
  & \textcolor{fillgauss}{G}  & 21.20 & .61 & .02 & 13.1 \\
\hline
\end{tabular*}
\end{table}

\section{Conclusion}

Masking fill and attribution method are not neutral experimental choices in perturbation-based XAI. Across three architectures and $1{,}100$ AudioSet clips, none of the three tested fills constitutes an information-absent baseline, with each activating a distinct class profile under full occlusion. Under partial masking, between 41 and 77\% of output displacement is orthogonal to the attribution axis. Logit-space geometric decomposition provides a principled black-box diagnostic for how fill and attribution method shape model outputs. For certain model--fill combinations, the off-axis component concentrates in a stable, low-dimensional subspace, affording mitigation opportunities that an isotropic residual would not. Evaluating any such correction, however, requires faithfulness metrics that themselves depend on fill choice. Understanding the causal mechanics behind the fill-specific output geometry requires opening the black box to examine intermediate activations.\looseness=-1

\newpage
\section{Compliance with Ethical Standards}
This is a computational study using the publicly available AudioSet dataset~\cite{Gemmeke2017}. No human subjects were involved and no ethical approval was required.

\section{Acknowledgments}

Nico Garc\'ia-Peguinho is supported by a doctoral studentship from the School of Electronic Engineering and Computer Science at Queen Mary University of London. David Kelly acknowledges support from the UKRI AI program and the Engineering and Physical Sciences Research Council for CHAI - Causality in Healthcare AI Hub [grant number EP/Y028856/1].

\bibliographystyle{IEEEbib}
\bibliography{strings,refs}

\begin{thebibliography}{10}

\bibitem{Jain2022}
Saachi Jain, Hadi Salman, Eric Wong, Pengchuan Zhang, Vibhav Vineet, Sai Vemprala, and Aleksander Madry,
\newblock ``Missingness bias in model debugging,''
\newblock in {\em Proc. ICLR}, 2022.

\bibitem{Qiu2022}
Luyu Qiu, Yi~Yang, Caleb~Chen Cao, Yueyuan Zheng, Hilary Ngai, Janet Hsiao, and Lei Chen,
\newblock ``Generating perturbation-based explanations with robustness to out-of-distribution data,''
\newblock in {\em Proc. ACM Web Conf.}, 2022, pp. 3594--3605.

\bibitem{Wickstrom2024}
Kristoffer~K. Wickstr{\o}m, Marina H{\"o}hne, and Anna Hedstr{\"o}m,
\newblock ``From flexibility to manipulation: {The} slippery slope of {XAI} evaluation,''
\newblock in {\em ECCV Workshop on Explainable Computer Vision}, 2024.

\bibitem{Rong2022}
Yao Rong, Tobias Leemann, Vadim Borisov, Gjergji Kasneci, and Enkelejda Kasneci,
\newblock ``A consistent and efficient evaluation strategy for attribution methods,''
\newblock in {\em Proc. ICML}, 2022, pp. 18770--18795.

\bibitem{Neely2021}
Michael Neely, Stefan~F. Schouten, Maurits J.~R. Bleeker, and Ana Lucic,
\newblock ``Order in the court: Explainable {AI} methods prone to disagreement,''
\newblock in {\em Proc. ICML}, 2021.

\bibitem{Krishna2022}
Satyapriya Krishna, Tessa Han, Alex Gu, Javin Pombra, Shahin Jabbari, Steven Wu, and Himabindu Lakkaraju,
\newblock ``The disagreement problem in explainable machine learning: A practitioner's perspective,''
\newblock {\em Transactions on Machine Learning Research}, 2024.

\bibitem{Lundberg2017}
Scott~M. Lundberg and Su-In Lee,
\newblock ``A unified approach to interpreting model predictions,''
\newblock in {\em Proc. NeurIPS}, 2017, pp. 4765--4774.

\bibitem{Ribeiro2016}
Marco~Tulio Ribeiro, Sameer Singh, and Carlos Guestrin,
\newblock ````why should {I} trust you?'': Explaining the predictions of any classifier,''
\newblock in {\em Proc. ACM SIGKDD}, 2016, pp. 1135--1144.

\bibitem{Petsiuk2018}
Vitali Petsiuk, Abir Das, and Kate Saenko,
\newblock ``{RISE}: Randomized input sampling for explanation of black-box models,''
\newblock in {\em Proc. BMVC}, 2018, p. 151.

\bibitem{Akman2024}
Alican Akman and Bj{\"o}rn~W. Schuller,
\newblock ``Audio explainable artificial intelligence: A review,''
\newblock {\em Intelligent Computing}, vol. 3, pp. 0074, 2024.

\bibitem{Parker2025}
James E.~K. Parker,
\newblock ``The planetization of machine listening,''
\newblock {\em Critical Inquiry}, vol. 52, no. 1, pp. 21--47, 2025.

\bibitem{Mishra2020}
Saumitra Mishra, Emmanouil Benetos, Bob L.~T. Sturm, and Simon Dixon,
\newblock ``Reliable local explanations for machine listening,''
\newblock in {\em Proc. IJCNN}, 2020, pp. 1--8.

\bibitem{Haunschmid2020}
Verena Haunschmid, Ethan Manilow, and Gerhard Widmer,
\newblock ``{audioLIME}: Listenable explanations using source separation,'' Proc. 13th Int. Workshop Mach. Learn. Music (MML 2020), 2020.

\bibitem{Wu2023}
Xiaoliang Wu, Peter Bell, and Ajitha Rajan,
\newblock ``Explanations for automatic speech recognition,''
\newblock in {\em Proc. ICASSP}, 2023, pp. 1--5.

\bibitem{Marinelli2024}
Luca Marinelli and Charalampos Saitis,
\newblock ``Explainable modeling of gender-targeting practices in toy advertising sound and music,''
\newblock in {\em Proc. ICASSPW}, 2024, pp. 818--822.

\bibitem{Hjuler2025}
Maja~J. Hjuler, Line~H. Clemmensen, and Sneha Das,
\newblock ``Exploring local interpretable model-agnostic explanations for speech emotion recognition with distribution-shift,''
\newblock in {\em Proc. ICASSP}, 2025, pp. 1--5.

\bibitem{Kelly2026a}
David~A. Kelly and Hana Chockler,
\newblock ``I guess that's why they call it the {Blues}: Causal analysis for audio classifiers,''
\newblock in {\em Proc. ACM Multimedia}, 2026,
\newblock to appear.

\bibitem{Kelly2026b}
David~A. Kelly and Hana Chockler,
\newblock ``If it's good enough for you, it's good enough for me: Transferability of audio sufficiencies across models,''
\newblock in {\em Proc. ACM Multimedia}, 2026,
\newblock to appear.

\bibitem{Sturm2014}
Bob~L. Sturm,
\newblock ``A simple method to determine if a music information retrieval system is a ``horse'',''
\newblock {\em IEEE Trans. Multimedia}, vol. 16, no. 6, pp. 1636--1644, 2014.

\bibitem{Green2024}
Owen Green, Bob L.~T. Sturm, Georgina Born, and Melanie Wald-Fuhrmann,
\newblock ``A critical survey of research in music genre recognition,''
\newblock in {\em Proc. ISMIR}, 2024, pp. 745--782.

\bibitem{Parekh2022}
Jayneel Parekh, Sanjeel Parekh, Pavlo Mozharovskyi, Florence d'Alch{\'e} Buc, and Ga{\"e}l Richard,
\newblock ``Listen to interpret: Post-hoc interpretability for audio networks with {NMF},''
\newblock in {\em Proc. NeurIPS}, 2022, pp. 35270--35283.

\bibitem{Paissan2024}
Francesco Paissan, Mirco Ravanelli, and Cem Subakan,
\newblock ``Listenable maps for audio classifiers,''
\newblock in {\em Proc. ICML}, 2024, pp. 39009--39021.

\bibitem{Fong2017}
Ruth~C. Fong and Andrea Vedaldi,
\newblock ``Interpretable explanations of black boxes by meaningful perturbation,''
\newblock in {\em Proc. ICCV}, 2017, pp. 3429--3437.

\bibitem{Covert2021}
Ian Covert, Scott Lundberg, and Su-In Lee,
\newblock ``Explaining by removing: A unified framework for model explanation,''
\newblock {\em JMLR}, vol. 22, no. 209, pp. 1--90, 2021.

\bibitem{Blucher2024}
Stefan Bl{\"u}cher, Johanna Vielhaben, and Nils Strodthoff,
\newblock ``Decoupling pixel flipping and occlusion strategy for consistent {XAI} benchmarks,''
\newblock {\em Transactions on Machine Learning Research}, 2024.

\bibitem{Sturmfels2020}
Pascal Sturmfels, Scott Lundberg, and Su-In Lee,
\newblock ``Visualizing the impact of feature attribution baselines,''
\newblock {\em Distill}, 2020.

\bibitem{Park2019}
Daniel~S. Park, William Chan, Yu~Zhang, Chung-Cheng Chiu, Barret Zoph, Ekin~D. Cubuk, and Quoc~V. Le,
\newblock ``{SpecAugment}: A simple data augmentation method for automatic speech recognition,''
\newblock in {\em Proc. Interspeech}, 2019, pp. 2613--2617.

\bibitem{Gemmeke2017}
Jort~F. Gemmeke, Daniel P.~W. Ellis, Dylan Freedman, Aren Jansen, Wade Lawrence, R.~Channing Moore, Manoj Plakal, and Marvin Ritter,
\newblock ``{Audio Set}: An ontology and human-labeled dataset for audio events,''
\newblock in {\em Proc. ICASSP}, 2017, pp. 776--780.

\bibitem{Kong2020}
Qiuqiang Kong, Yin Cao, Turab Iqbal, Yuxuan Wang, Wenwu Wang, and Mark~D. Plumbley,
\newblock ``{PANNs}: Large-scale pretrained audio neural networks for audio pattern recognition,''
\newblock {\em IEEE/ACM Trans. Audio, Speech, Lang. Process.}, vol. 28, pp. 2880--2894, 2020.

\bibitem{Gong2021}
Yuan Gong, Yu-An Chung, and James Glass,
\newblock ``{AST}: Audio spectrogram transformer,''
\newblock in {\em Proc. Interspeech}, 2021, pp. 571--575.

\bibitem{Welford1962}
B.~P. Welford,
\newblock ``Note on a method for calculating corrected sums of squares and products,''
\newblock {\em Technometrics}, vol. 4, no. 3, pp. 419--420, 1962.

\end{thebibliography}

\end{document}